# Radio Frequency Identification: Decades at a Time

Christopher Saetia, *Graduate Student Member, IEEE*, Daniel M. Dobkin, and Gregory Durgin, *Fellow, IEEE*

*Abstract*— In this article, we briefly review the history of the use of radio signals to identify objects, and of the key Radio Frequency Identification (RFID) standards for ultra-high-frequency (UHF) and near-field communications that enabled broad use of these technologies in daily life. We will compare the vision for the future presented by the Auto-ID Lab in the early 21$^{st}$ century with the reality we see today, two decades and a little after. We will review some of the applications in which UHF RFID technology has become hugely successful, others where High Frequency Near-field Communications (HF NFC) is preferred, and applications where optical identification or active wireless communications are dominant.

We will then examine some possible future paths for RFID technology. We anticipate that UHF read capability will become widely available for cellphones, making it as universal as NFC and Bluetooth are today. We will look at more sophisticated radio interfaces, such as multiple-antenna phased arrays for readers, and tunnel diode reflection for tags. We will discuss the integration of information from Artificial Intelligence (AI)-based image processing, barcodes, NFC and UHF tags, into a digital twin of the real environment experienced by the human user. We will examine the role of RFID with sensing in improving the management of perishable goods. The role that RFID might play in a truly circular economy, with intelligent recycling and reuse, will be discussed. Finally, we survey the many hazards and obstacles that obstruct the path to an RF-informed future.

## I. INTRODUCTION

F OR most of the existence of humans on earth, objects have been identified by their visual appearance, with the addition of unique visible markings in recent millennia. With the invention of radio communications in the late 19$^{th}$ century, alternative means of identification became possible, though decades passed before the first uses of the technology for this purpose were encountered. Early researchers investigated the use of both active and passive devices – that is, devices that transmit their own signals, and devices that merely modulate reflected signals from another transmitter. Both approaches have been greatly elaborated over the decades, and both have found wide applications in the modern world. In this paper, as summarized in Figure 1, we first review these early investigations and the resulting visions for a radio-identified world. We then compare those visions to where we find ourselves today, and (hopefully enlightened by this investigation of our history) try to project what might happen over the next few decades of radio identification.

## II. THE HISTORY OF RADIO FREQUENCY IDENTIFICATION

### A. Early Developments in Radio Detection and Identification

The fundamental economics of identification are that you cannot spend more to identify an object than what that identification is worth. Radio technology began in the late 19$^{th}$ century, and commercial radio communications technologies were in wide use by the second decade of the 20$^{th}$ century, but the use of radio to identify an object had to wait until radio was first used to detect an object: that is, until the invention of RAdio Detection And Ranging (RADAR). Once an airplane was detected by a RADAR system, it was important to figure out whether it was an enemy bomber or simply a friendly fighter returning to base – a technology known as Identification, Friend or Foe (IFF) [1]. Early work focused on transponders, active radios mounted in an airplane that detect an inquiry signal at one frequency (here 1030 MHz) and transmit an identifying reply, typically on a different frequency (1090 MHz).

Transponders are still widely used in ordinary civilian aircraft identification as well as military applications, but even in the modern world they cost US$2,000 or more. This high price of transponders is fine for finding aircraft, whose cost is hundreds of thousands to hundreds of millions of dollars, and where the consequences of misidentification can involve loss of many human lives. To employ radio signals to identify smaller, less expensive objects than airplanes, it was necessary to reduce the size, complexity, and cost of the identifying equipment. This economic requirement could be satisfied both through passive reflective devices, and through lower-cost active devices.

One of the early successful implementations of passive radio identification was the rail industry in the United States. The United States Bureau of Transportation Statistics reports that there were about 1.7 million freight cars in the United States in 2023 [2], and the U.S. freight network traverses about 140,000 route miles (225,308 km) [3]. The rail industry has always needed to track many expensive cars over lengthy routes. Further, the known routing of the rails means that simply reading at appropriate points provides accurate location information [4]. By the early 1990's, passive transponders following the Association of American Railroads (AAR) S-9018 standard at 902-928 MHz, were coming into wide use (Figure 2). On the other hand, the automobile industry is more diverse, both in types of vehicles and in the administrative complexity of varying state laws (in the US), and the problem of tracking is more challenging since roads are bigger than rails.

Daniel M. Dobkin is with Enigmatics, 877 Sutter Ave., Sunnyvale, California (email: dan@enigmatic-consulting.com). Christopher Saetia and Gregory Durgin are with the Georgia Institute of Technology, North Avenue, Atlanta, Georgia (emails: csaetia3@gatech.edu, durgin@gatech.edu).

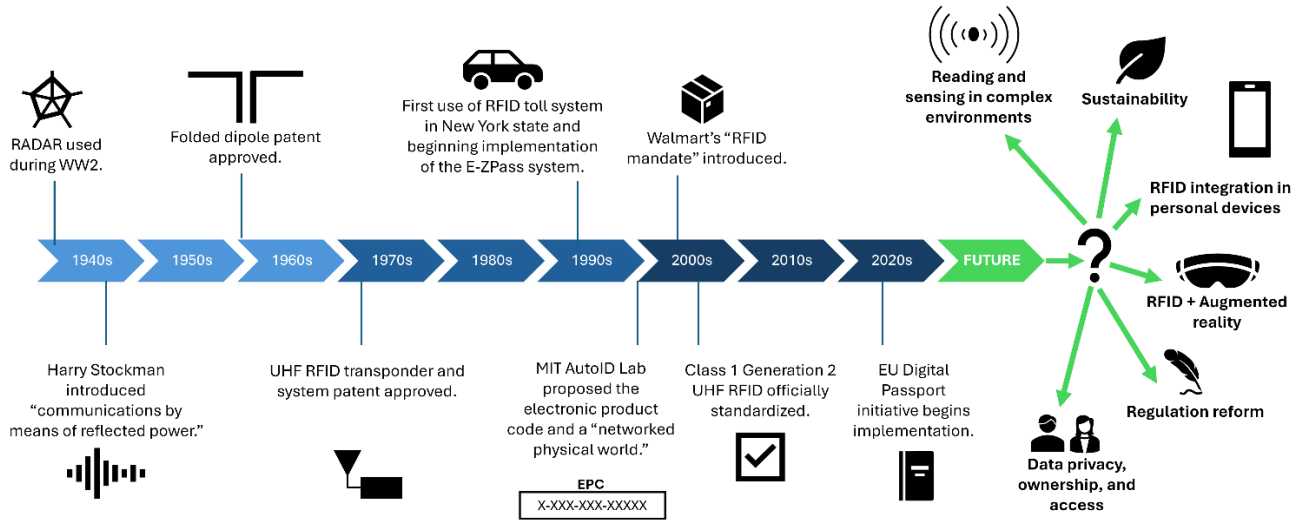

**Fig. 1.** The use of RF for identification originated from the use of RADARs during World War II. For traffic tolling, Mario Cardullo invented the first UHF RFID transponder design and was granted a patent in 1973 [5]. The 1990s saw the growing use of RFID in America's transportation system, such as the inter-state E-ZPass toll system debuting in 1993 [6] Starting in the 2000s, RFID technology has spread for identifying retail inventory, exposing every-day people to RFID. The future of RFID in the next decades involves further integration of RFID in cyber-physical systems, leading to complex issues that need to be solved, such as ethical data management, operation in complex environments, etc.

Various radio approaches, such as E-ZPass and Title 21, have been used in differing jurisdictions, although much of the world is migrating to interoperable ISO 18000-6C RFID for open road tolling, from the local, ad-hoc, or proprietary systems that dominated open tolling applications. In the 1990's, technologies for tracking shipping containers and other large objects were standardized under ANSI 371.1 and 371.2, using battery-powered active transmitters [7].

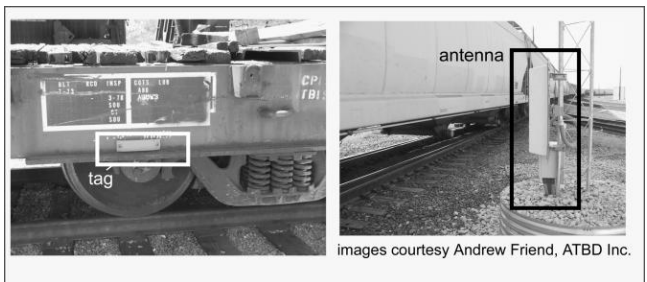

**Fig. 2.** An example of a typical passive tag on a train undercarriage (left) and a fixed, rail-side reader antenna (right) for identification of railcars [8].

In contrast to the far-field approaches described above, where radiating signals are employed, it is also possible to employ inductive coupling between the transmitter and receiver when they are close to one another. This approach has been standardized in ISO 11784/85 and 14223, and is widely used for identification of cows and other farm animals, where the associated frequencies of operation (in the hundreds of kHz) are not strongly affected by the moisture in a cow's body, and tracking at short range (a meter or less) can be used [8]. The cost of the transponder, on the order of US$1, is insignificant for a cow, although a challenge for smaller animals such as chickens. Similar tracking of people (or at least their bank accounts) using short-range inductively-coupled tags was standardized as ISO 14443 and 15693, requiring the addition of a silicon chip and antenna wiring into plastic cards used for credit payment and related purposes.

By the 1990's, integrated circuit (IC) technology had advanced sufficiently to implement simple communications protocols and memory on a single, inexpensive chip. The potential of the technology for identification had been demonstrated in proprietary technologies such as the Texas Instruments Tag-IT and Philips u-code products, for specialized applications such as tracking library books [9]. Improvements were also being applied to low-cost, low-power active radio links, with the creation of the Bluetooth standard (later IEEE 802.15.1) in the late 1990's.

*B. The Auto-ID Lab at the Massachusetts Institute of Technology*

A broader vision for radio identification was introduced by MIT researchers David Brock, Sanjay Sarma, Sonny Siu, Kevin Ashton, and Eric Nygren around 1999, who proposed the use of a globally unique Electronic Product Code (EPC) to identify every manufactured object, and a software infrastructure to provide access to information about the object so identified [10][11]. They envisioned a world in which each manufactured item was tagged and uniquely identified, supporting on the order of $10^{12}$ unique objects each year. A single open architecture would allow identification of every networked physical object, covering both very low-cost and more elaborate tag approaches. The EPC would resemble the universal product code (UPC) then used for bar code identification, but would add the capability to identify a unique object rather than only a type of object. The initial concept for the EPC proposed a 96-bit identifier, potentially allowing about $10^{29}$ uniquely identified objects, or roughly $10^{19}$ objects per person for a population of ten billion humans (although allocation of bits to specific applications, such as the manufacturer or product ID, reduces the number of objects that can be uniquely identified). They envisioned millions of readers on floors, trucks, shelves and other locations, with automated reading replacing handheld readers operated by humans. The readers would be networked using the standard internet protocols (TCP/IP), and data would be available through a unique Object Naming Service (ONS), which would provide universal access to every physical object, just as the Internet Naming Service (INS) provides access to every node on the internet. The availability of a unique identification could benefit every aspect of the supply chain [12], as shown schematically in Figure 3.

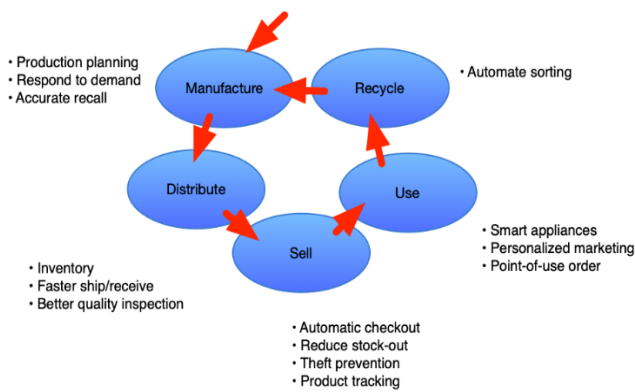

**Fig. 3.** Application of unique identification and its benefits when a product or component enters the fully circular supply chain [12].

The Auto-ID work initially considered four candidate industrial-scientific-medicine (ISM) frequencies (125 kHz, 13.56 MHz, 900 MHz, and 2.45 GHz) for use in identification, but most actual protocol and hardware development focused on the circa-900 MHz band (with the actual frequencies dependent on regulatory jurisdiction). The Center envisioned as many as five classes of tags, ranging from Class 0/1 read-only passive tags to Class V tags [13], which would be regarded in modern parlance as mesh-networked readers. Two early standards, Class 0 and Class 1, were implemented in cooperation with sponsor companies, and commercial tags using these standards were sold in significant quantities until replaced by Class 1 Generation 2. The Class 0 and Class 1 standards used similar reader symbols, but were otherwise completely incompatible, with differing tag symbols, medium allocation, and collision detection schemes [14]. (A third incompatible approach, standardized as ISO18000-6B, was also implemented outside the activities of the Auto-ID Center.)

The Class 1 Generation 2 (C1G2) standard, already envisioned in the 2002 discussions [13], was completed in 2004, and was approved as ISO 18000-6C in 2006. The C1G2 standard addressed many of the problems encountered in the early standards: binary-tree collision resolution was replaced with an ALOHA-based random collision avoidance approach that allowed tags without unique EPC's to participate, simple link-level security was added, and explicit specifications for memory maps, programming, and locking were included.

In 2004, roughly coincidental with the completion of the C1G2 standard, Yossi Sheffi of the MIT Center for Transportation and Logistics published a paper on the status of RFID innovation [15]. He presented RFID as intended to replace bar codes to track items in the supply chain, envisioning a future where a low-cost tag is used to provide a unique identification to every item and readers are networked to databases that store information about the item. He represented the "cycle of innovation" as being composed of six stages that he phrased as: "the fog of innovation", "life support for the existing", "the stamp of approval", "out with the old", "it's everywhere," and "the big bang." The paper provided examples for each of these stages for the development of five existing technologies: refrigeration, automobiles, electric lighting, television, and personal computers (although not all the stages for each technology are very well-defined). In this context, he claimed that RFID was in the early stages, but possibly had reached the "stamp of approval" level with support from Wal-Mart, Sun Microsystems, the US Department of Defense, Gillette, and other commercial companies. He also mentioned that passive low-cost RFID might interact with Zigbee and WiMax active technologies, although Bluetooth was not included.

In the next section we compare what people envisioned to what has happened in radio identification.

III. THE RADIO IDENTIFICATION WORLD TODAY

Radio-based identification today is widely used, but is not based on any single technology or protocol. Instead, each technology has found an optimal niche or niches, and substantial markets still employ vision-based recognition, which has also seen tremendous innovation. In this section we review the spectrum used for RFID and its various applications. Then, we further examine the various identification applications, such as the bar and QR code identification, 13.56 MHz NFC, passive UHF RFID, and Bluetooth and other active beacon technologies, and the applications in which each differing approach is dominant.

*A. The Spectrum of RFID*

RFID applications must carefully plan and choose their operation in an increasingly crowded radio spectrum. The chart in Table 1 provides a rough guide to typical RF band nomenclature in engineering that has also been adopted widely by the RFID community. There are numerous standards, some of which conflict in common usage. Originally, most RF bands were designated by wavelength in free space, referenced

### ITU Band Designations

| Common Usage | Frequency Range | Wavelength Range | Radio Frequency Designation |
|---|---|---|---|
| HF | 3–30 MHz | 10-100 m | **H**igh **F**req |
| VHF | 30–300 MHz | 1-10 m | **V**ery **H**igh **F**req |
| UHF | 0.3 – 3 GHz | 10 cm - 1 m | **U**ltra **H**igh **F**req |
| ////// | | | |
| ////// | | | |
| micro-waves | 3–30 GHz | 1 cm - 10 cm | **S**uper **H**igh **F**req |
| ////// | | | |
| ////// | | | |
| mm-waves | 30–300 GHz | 1 mm - 1 cm | **E**xtremely **H**igh **F**req |
| ////// | | | |
| Tera-Hertz | > 300 GHz | < 1 mm | **T**remendously **H**igh **F**req |

### IEEE Band Designations

| Range | Band | Legacy Name |
|---|---|---|
| 0.3-1 GHz | UHF | **U**ltra **H**igh **F**req |
| 1-2 GHz | L | **L**ong wave |
| 2-4 GHz | S | **S**hort wave |
| 4-8 GHz | C | **C**ompromise |
| 8-12 GHz | X | Cross (**X**) band |
| 12-18 GHz | Ku | **K**urz-**u**nder |
| 18-27 GHz | K | **K**urz* |
| 27-40 GHz | Ka | **K**urz-**a**bove |
| 40-75 GHz | V | **V**ery high |
| 75-110 GHz | W | **W** after V |
| 110-300 GHz | G | **G**reater |

* "kurz" is German for "short"

////// overlapping or conflicted nomenclature in common use by decade. For example, low-frequency (or simply LF), referred to any radio wave frequency with a corresponding wavelength between 100 m and 1 km. This practice has been standardized by the International Telecommunications Union (ITU) [16].

**Table 1.** Chart surveying the common nomenclature for RF bands according to International Telecommunication Union and Institute of Electrical and Electronics Engineers standards.

In the upper UHF, microwave, and millimeter wave region of radio spectrum, however, engineers often use a lettering system with tangled historical origins to designate bands -- a practice that has been standardized by the Institute for Electronic and Electrical Engineers (IEEE) [17]. Overlaid upon this nomenclature, the higher-frequency bands are often generically referred to as "microwave", "mm-wave", and "TeraHertz" bands, with usage often blurring the wavelength decade boundaries. Complicating the picture further is the usage of a different lettering system to refer to RF bands that is borrowed from the lexicon of standardized waveguide components of different frequency ranges; the waveguide lettering convention is not aligned (and is sometimes contradictory) with the IEEE microwave band designations.

Table 2 is a listing of radio frequency bands that are officially US-designated *industrial-scientific-medical (ISM)* bands [18], [19]. Transmitters in these bands must satisfy the Federal Communications Commission's (FCC's) Section 47, Chapter 1.A, Part 15 Rules for unlicensed RF Devices [20]. These rules specify transmit power maxima, bandwidth limitations, signal types, antenna types, and other operational details that allow RF devices to transmit without a license without overwhelming the shared spectrum resource.

The bands designated *ISM* are not the *only* RF bands that allow unlicensed radio transmissions. For example, *Citizen's Broadband Radio Spectrum (CBRS)* from 3550-3700 MHz, allows unlicensed transmissions in areas that other *incumbent* and *licensed* users have not excluded [21]. But the ISM bands are in the same family of commonly used spectrum for low-cost radios. Although Table 2 reflects the US designations, the US and most of the world have mutually aligned their unlicensed bands so that these designations reflect a near-consensus worldwide allocation of spectrum with only minor deviations.

Because of the ad-hoc nature of RFID applications – low-cost, low-powered devices with sporadic usage across highly-varied environments – most commercial RFID applications fall within ISM spectrum designations, with the most common being the 13.5 MHz HF and 915 MHz UHF designations, and the 2.4 GHz S-band.

| Start (MHz) | Stop (MHz) | Bandwidth (MHz) | Band Name | Common Uses |
|---|---|---|---|---|
| 6.765 | 6.795 | 0.030 | HF | RF heating; *wireless power;* tissue heating; underwater comms |
| 13.553 | 13.567 | 0.014 | HF | **RFID;** *near-field communications/electronic payments;* inductive heating |
| 26.957 | 27.283 | 0.326 | HF | citizens band radio; RC toys |
| 40.660 | 40.700 | 0.040 | VHF | RC vehicles; *garage door openers;* baby monitors |
| 433.050 | 434.790 | 0.740 | UHF | *garage door openers;* wireless sensors; RC vehicles |
| 902 | 928 | 26 | UHF | **RFID;** short-range comms; wireless sensors; *utility meter reading* |
| 2400 | 2500 | 100 | microwave S-band | **RFID;** wireless networking; Wi-Fi; Bluetooth; baby monitors; wireless microphones |
| 5725 | 5875 | 150 | microwave C-band | wireless networking; Wi-Fi; fixed wireless networks; drone video |
| 24,000 | 24,250 | 250 | microwave K-band | amateur radio; amateur satellite; radars; tissue heating |
| 61,000 | 61,500 | 500 | microwave V-band | high-throughput, short-range communications; *vehicle radars;* vital sign monitoring |
| 122,000 | 123,000 | 1,000 | microwave G-band | *short-range radar;* amateur radio |
| 244,000 | 246,000 | 2,000 | microwave G-band | *radar and imaging;* non-destructive testing |

**Table 2.** Industrial-Scientific-Medical (ISM) bands in the United States and the common applications that use them.

These primary RFID applications are listed in **bold** in Table 2. The 13.5 MHz HF band RFID systems use inductive coupling and have short-by-design read ranges for applications that require secure, unambiguous transactions (transit farecards, electronic payments, limited-access entry, etc.) The 915 MHz UHF band RFID systems rely on far-field transmission and are suited for applications where long-range, non-proximity reading is essential (warehouse logistics, retail inventory, open highway tolling, stand-off measurement, etc.) There are also a number of other *italicized common uses* in Table 2 that reflect forms of radiofrequency identification that may not commonly bear the moniker "RFID" in practice but perform similar low-power wireless identification.

Having reviewed the various radio bands available for use, let us now see each technology uses radio (or other) signals to identify objects in today's real world.

*B. Bar and QR Codes*

Bar codes, once read by a laser beam and today mostly read using imaging, remain the dominant automated means of identifying an object in the retail world, with roughly one billion products identified and 10 billion scans per day in 2024 [22]. As shown in Figure 4, a typical linear bar code can have 8 to 12 characters, and contains a product type, but not a lot number or unique product identification. Various larger bar codes are also used [23]. Standards support a wide variety of QR codes, with up to 177x177 readable features in addition to finders, timing, and alignment [24]. These codes are widely used for accessing web sites as well as for identifying items. Unique QR codes can be printed onto the label of an object at modest cost, and the international standards body GS1 is advocating that all retailers be equipped to read QR codes as well as bar codes by 2027. The majority of wireless phones can use their cameras to read and decode QR codes, enabling the codes to be useful both to retailers and customers. Automated visual identification has substantially replaced human visual identification in retail, but visual identification remains the single dominant product identification technology in retail and trade.

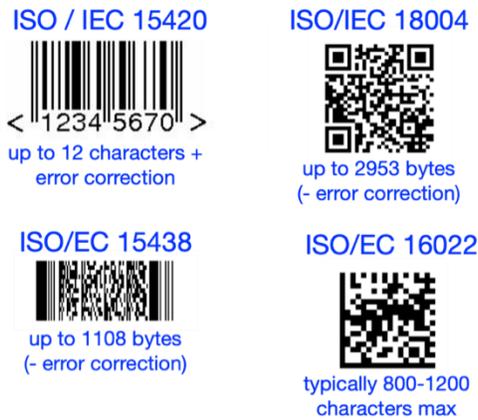

**Fig. 4.** Examples of bar and QR codes.

*B. Near-Field Communications*

Tags employing near-field communications, primarily following the 13.56 MHz ISO 14443 standard, are widely used for identification of humans in the context of building or facility entry, and as tickets for temporary access (Figure 5). They are also used for transit passes and other entry-control applications. Tags can also be used to identify pets and livestock, although this often employs lower-frequency technologies. A reader includes a powered inductive coil typically a few centimeters on a side. A tag with a few-turn inductive antenna is incorporated into a plastic card roughly 5x9 cm. These are very commonly used to identify assets of a given person, such as a bank account, in payment for purchases of goods or service.

Short range is a key to applications for this technology. The additional power available when the passive device is only a few cm from the reader makes it possible to employ encryption technologies to improve the security of transactions. It is not desirable for a person's credit cards to be readable except when they are intentionally placed near a reader to make a payment. Similarly, a mechanism to control entry to a room or building should read only the entry control mechanism presented to it. Current sales of HF NFC tags and cards of various types appear to exceed 10 billion per year in 2024.

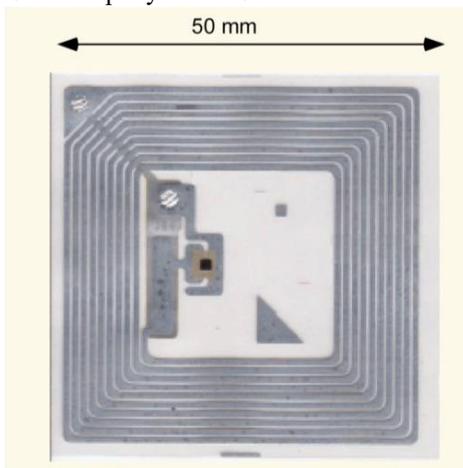

**Fig. 5.** Example of a 13.56 MHz inductive tag.

*C. Far-Field RFID*

The ISO 18000-63 standard, which is based on the EPC Class 1 Generation 2 standard and the earlier ISO 18000-6C standard, is the most used approach to identification of objects for inventory management and retail sale using far-field radio techniques. The commercial deployment of the technology has been greatly aided by the formation of the RAIN alliance in 2014 to advocate for use and standardization of UHF RFID technology. Practical implementation has also benefited from test services provided by such organizations as the University of Auburn ARC Laboratory, providing aid in the optimal selection and placement of tags on a wide variety of products. A number of large retail vendors, such as Uniqlo, Victoria's Secret, and Lululemon, use EPC compliant UHF tags on most of their retail items (often in conjunction with bar and/or QR codes, as shown in Figure 6). Also, automated checkout is employed by Uniqlo and others. Wal-Mart has also expanded its use of tagged items.

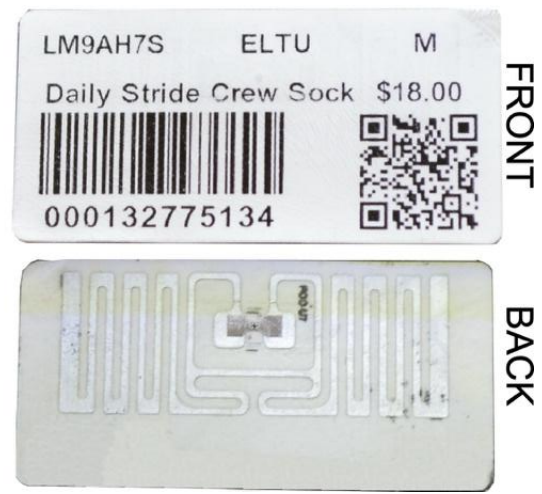

**Fig. 6.** Example of far-field UHF RFID tag with bar code and QR code labels.

The use of far-field RFID in retail provides substantial benefits in inventory management as well. High-quality item counts are readily performed in stores, so that inventory can be taken monthly or weekly, where manual count of bar codes is often performed only on a yearly or biannual basis.

As a result of this long-term commercial development, 52.8 billion RAIN RFID tag chips were shipped worldwide in 2024 [25]. While this is a substantial growth over the last two decades, it still represents a small fraction of the overall retail identification market, as is apparent by comparison with Section III.B above.

*C. Active Beacons*

A number of active-beacon technologies, in which a radio attached to an object or human transmits an identifying signal at a regular interval or in response to a query or event, are in wide use. The most common technologies are Bluetooth Low Energy (BLE), which is based on the IEEE 802.15.1 standard, and IEEE 802.15.4-based communications, including Zigbee and Matter standards. BLE operates in the 2.4 GHz band that is available in most of the world. It is generally like the short-range Bluetooth standard, but is modified to support reduced energy use, with a simplified protocol stack and data rates as low as 125 kbps for long-range use, though most communications are at 1 Mbps [26]. Most beacons simply broadcast at fixed time intervals in an advertising mode.

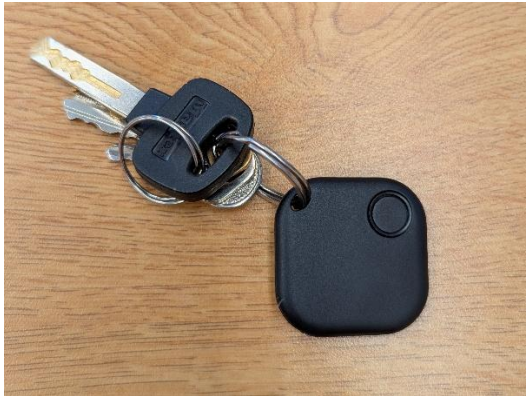

**Fig. 7.** Example of a commercially available Bluetooth beacon used for tracking.

BLE beacons are used in automotive applications (such as keys or sensors), proximity monitoring and advertising, asset tracking and location, patient tracking for healthcare, and numerous other applications. In some cases, the beacons are co-packaged with ultrawideband transmitters, using IEEE 802.15.4z, to provide very precise short-range location services. On the order of 80 million Bluetooth tracking devices are expected to be sold in 2025.

*D. Multiple Standards, Multiple Uses*

As is apparent from the discussion above, no one standard or approach is universally used for automated identification of physical objects. Instead, each identification technology tends to grow in the specific application context where it provides the right mixture of capability and cost. Visual identification is extremely inexpensive, generalizable to almost any smart device that has a camera, and with the addition of QR codes has substantial capacity for unique information related to the object to which the identifier is attached. Inductively-coupled HF devices are intrinsically short-range (on a human scale) and thus allow their utilization to be limited to their intended purposes by the associated human. Inductive transmitters and receivers are often integrated into or easily added to handheld phones, and thus allow the phone to both read and emulate passive HF devices. Radiatively-coupled UHF tags supplement visible codes and provide huge improvements in inventory management, but are more expensive than visible coding, particularly when the associated object is electrically conductive. Active microwave radio beacons are substantially more expensive than any passive device, but provide communications services to a wide variety of receivers, and can support time-of-flight-based location that is difficult to accomplish with passive devices.

Thus, in Sheffi's terms, passive UHF RFID remains in the "life support for the existing" stage: alternative technologies have prospered even as passive UHF RFID has expanded tremendously. The identification of the physical world, even specifically through radio means, remains multifaceted, with different approaches optimal for different problems.

IV. THE NEXT DECADE OF RFID DEVELOPMENT

In this section we will discuss the next set of likely innovations, including those taking place today or expected to become significant over the next few years.

*A. UHF RFID in Phones*

We expect UHF RFID to become available on conventional cellular phones within a couple of years, leading to much wider use of UHF RFID in retail, asset tracking, and related applications. Handsets with UHF RFID are already available from specialty vendors [27]. The RAIN Alliance has released statements from Qualcomm, Impinj, and Decathlon suggesting that these vendors anticipate wide availability of mobile phone handsets with UHF RFID capability over the next few years. The availability of readers on phones will greatly reduce the cost of implementation in retail and enable the wide use of UHF RFID in smaller businesses, where it is not practical today. As noted above, the benefits of such use include huge improvements in retail inventory frequency and accuracy. The time required to find items for online orders is also reduced substantially in many cases.

Consumer use of UHF RFID may also grow, although as noted above, QR codes provide similar functional advantages, and a short-range reading mode will be needed for consumers accustomed to reading what they are looking at, not what is on a shelf several meters away. On the other hand, the benefit of short read range in current HF RFID applications still applies, and it is not apparent that UHF RFID will replace HF RFID in identification-related uses.

*B. Reading in Complex Environments*

As noted above, annual sales of UHF RFID tags today are in the tens of billions. Larger numbers of RFID tags will be observed in many environments, both intended and unintended. Dealing with such large numbers of tags in complex environments may benefit from the use of more sophisticated radio configurations, such as phased-array multiple-in multiple-out (MIMO) reading approaches. The availability of multiple antennas enables beamforming to be used to see multiple regions; it has been demonstrated that large areas can be read with near-100% effectiveness using large numbers of phase configurations, as shown in Figure 8 below [28][29]. With phased arrays, multiple tags can be read simultaneously [30].

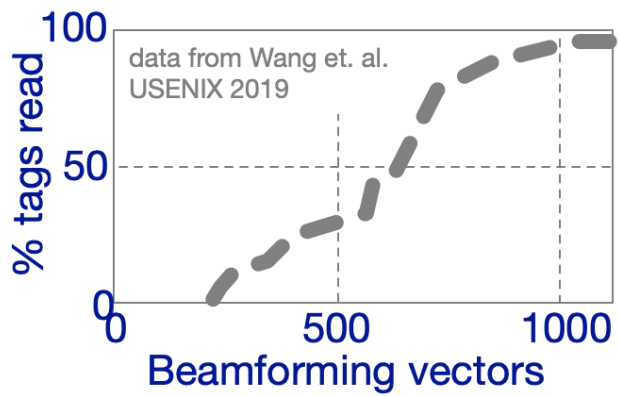

**Fig. 8.** Use of multiple synchronized antennas and beamforming to increase read percentage over a large area (data from [28]).

We believe phased array reading techniques will see increasing use in high-volume read environments, although local permission and/or regulatory changes may be needed in some cases. Current regulations for unlicensed radio operation in the 902-928 MHz band in the United States require that independent transmitters be uncoordinated. Site-specific licenses may be required for phased-array transmission. The phase-coherent addition of multiple transmission sites adds potentials or fields rather than power, and thus the total power can scale as the square of the number of transmitters. This scenario may lead to much higher local power density than is typically encountered in unlicensed applications, and safety issues should be examined before widespread use is encouraged.

A different approach to dealing with complex environments is to create localized links which collectively span a larger area. In the context of passive RFID, a possible method for accomplishing such links is the use of tag-to-tag communications. Nikitin et. al. reported passive communications using ambient backscatter over short distances of a few centimeters in 2012 [31]. A simplified version of this approach, in which multiple tags simply act together to talk to a distant reader, was demonstrated by Dobkin et al. [32] to increase backscattered power by a factor greater than five at range of 3 meters. Extensive research on the use of tag-to-tag communications has been reported in recent years; some representative work can be found in references [33], [34], [35]. The fundamental advantage of tag-to-tag communication is the potential for long-distance communication using multiple short-distance links. The fundamental challenges are the need for a reliable source of ambient radiation to backscatter, and the limited capabilities of a processor powered by ambient radiation. We speculate that the key long-term utility of tag-to-tag communications will be in the area of large-area sensor networks for e.g. mechanical structural integrity, where speed of communication is not important, but the capacity for multi-year and even multi-decade battery-free operation is critical.

Additionally, higher-frequency usages of RFID have been proposed and studied, such as 5.8 GHz C-band and mm-wave frequencies [36], [37]. These technologies are promising for precise localization, extended passive ranges, higher throughputs, and more sensing and measurement capabilities [38], [39], [40], [41]. Ultimately, though, bands above UHF become increasingly dependent on establishing a line-of-sight between RF reader and tag – and therefore also compete with optical identification and measurement technologies.

*C. Harmonic Detection and Nonlinear Signal Processes*

Clutter, multi-path, and strong self-interference at the RFID reader provide challenges to being able to detect RFID tags and sense their backscattered message. Hence, *harmonic RFID* has been proposed as a solution to mitigate the effects of interference from the reader on detecting and reading tags (Figure 9). Harmonic passive RFID tags can place identifying markers or information on frequencies away from the carrier frequency, with such possible frequencies being the carrier's harmonic frequencies or harmonics around the carrier frequency. The non-linearity of these tags' components allows them to produce these harmonic components and thus be detected by the readers despite self-interference.

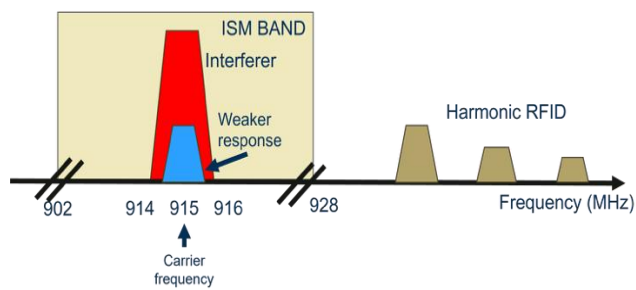

**Fig. 9.** Reflected signals at the fundamental carrier frequency, shown in blue, may encounter interference from the source of the signals being reflected, shown in red. Harmonic signals, shown in gold, may avoid interference from the transmitter.

An early instance of harmonic RFID was demonstrated by Staras, Klensch, and Rosen, who described a harmonic RFID tag meant to help identify vehicles [42]. The tag used a diode, a non-linear component, to generate a second harmonic that allowed the vehicle's ID to be read by a nearby receiver.

As demonstrated by Barbot [43], the non-linear electronic components in the tag's chip can produce detectable harmonics around the carrier frequency, even if the power at the tag is lower than the chip's sensitivity. Kumar et al. [44] used a standard UHF RFID tag integrated-circuit (IC) with a wide-band antenna, and a standard RFID reader with a custom down-conversion front-end, to read the third harmonic produced by the tag IC, demonstrating the integration of harmonic RFID with regular RFID systems. It seems possible that such harmonic detection might be combined with the phased arrays described in Section IV.B above, to enable detection of the presence of tags whose received power is insufficient to fully power their IC, and adjust phase to increase their received power so they can be read, as shown schematically in Figure 10 below.

Likewise, the harmonics of these tags can be leveraged to identify specific tags. Piumwardane et al. [45] varied duty-cycles of the baseband messages that are mixed with the carrier/interrogating signals to create distinct harmonic signatures for each tag. Saetia et al. [46] proposed the use of tunnel diodes (discussed in more detail in Section IV.D below) to produce distinct wide-band and harmonic signatures, when operating as oscillators, to be used for natural identification of tags. Harmonic RFID has promise for food monitoring [47], cough detection [48], temperature sensing [49], and other tracking and sensing applications.

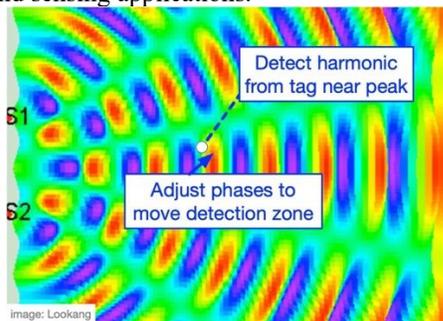

**Fig. 10.** Detection of harmonic signals from two readers S1 and S2, to enable optimization of phased array reader coverage (based on image from Wikimedia).

Harmonic RFID faces several challenges to full implementation. Harmonic systems typically require wide-band or dual-band antennas and receiver design, and low-conversion-loss designs for generating the harmonics. Harmonics can fall outside of the industrial-scientific-medicine (ISM) band and into bands licensed for other uses, thus creating the possibility of regulatory obstacles to wide use. Section V.A below briefly examines the possibility of regulatory reforms that might enable more flexible use of currently restricted frequencies.

### D. Tunnel Diodes and Backscatter Modulator Re-design

Path loss from the round-trip communication link between reader and tag significantly degrades the strength of the received backscattered signal at the reader, compared to traditional independent one-way links, and limits communication range from tag to reader as tags do not have traditional amplifiers to amplify the backscattered signal strength. Tunnel diodes are referred to often as "negative resistance" devices, because they have a negative slope in the current-voltage (I-V) curve (Figure 11). These diodes are used to make reflection amplifiers, because the negative resistance allows the reflection coefficient, which is always less than 1 in magnitude for a conventional passive load, to exceed unity for a tunnel diode. Many tunnel diodes require only modest bias voltages (as low as 69 mV [50]) to achieve negative resistance, and thus can consume minimal DC power (as low as 20 μW [53]).

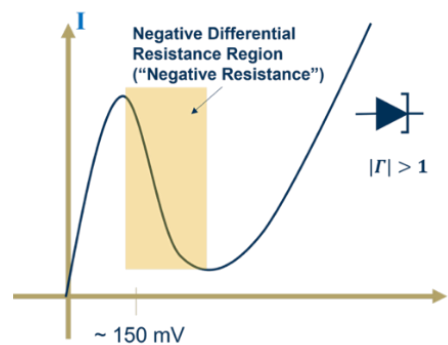

**Fig. 11.** Tunnel diode current-voltage curve, showing region of negative differential resistance.

The possibility of reflection amplification and low power consumption has led to research on the practical use of tunnel diodes in RFID. Amato et al. [50] demonstrated the use of tunnel diodes to backscatter a Manchester encoded message at a reader-tag range of 23 meters, while consuming only 29 μW of power. In later work, Amato et al. [51] demonstrated the capabilities of tunnel diodes to create detectable backscatter signals at 1 kilometer range. Saetia and Durgin [52] demonstrated that tunnel diode backscatter modulators, when biased at different voltages, can generate multi-bit symbol constellations for scalable higher-order modulation.

Tunnel diodes have been used for functionalities other than reflection amplification, to improve communication links. Eid et al. [53] utilized a tunnel diode, not only for reflection amplification, but also as an oscillator at 5.8 GHz, for low-power signal generation. Similarly, Varshney et al. [54] have worked on designing a single tunnel diode circuit that can switch between backscattering and producing its own signal for transmission in scenarios where a backscatter node receives no interrogating signal.

The low-power consumption of tunnel diodes makes them ideal candidates for future backscatter modulator designs. The main issue with utilizing tunnel diodes in commercial products today is the lack of manufacturing and commercial availability of the diodes. However, recent research has demonstrated several promising approaches that might allow integration of tunnel diode fabrication into a standard silicon process. For example, Zhu et al. have shown that near-monolayer films of $MoS_2$ can be lifted from a sapphire substrate and mechanically transferred to patterned silicon to create tunnel diodes with negative differential voltage regions, although the required applied voltage is still relatively high (around 3 volts) [55]. Tunnel diodes have been fabricated in silicon nanowires, although room-temperature negative resistance operation would be required [56].

### E. Sensing and Sustainability

Integration of passive tags with sensing is expected to become more common, particularly in the monitoring of perishable foods and other commodities [57]. The EU's development of digital passports is consistent with incorporation of either or both of UHF and HF RFID technologies on many goods, to provide detailed identification and characterization over the full life cycle of products. The use of UHF RFID tags to sense the concentration of chemicals such as oxygen, $CO_2$, and ethanol, as well as temperature, humidity, and internal pressure, has been reported [58]. Tags can also be electrically integrated with low-power sensors to detect ammonia and potassium.

Tag chips have very small volume, and are mainly composed of silicon, which in the environment will be oxidized to silicon dioxide, one of the common components of the Earth. Tag antennas have been traditionally formed using metals which might have more significant environmental impacts, so research has been reported on polymer conductors for tag antennas [59]. Work has been reported on the use of sewn conductive fibers integrated into clothing for sustainable tracking [60], [61]. Such technologies will help support regulatory requirements such as the EU Digital Product Passport and California SV707 requirements [62]. Tag backscatter responses may also be useful to help classify the materials to which they are attached [63]. However, the biggest medium-term benefits may simply be improvements in clothing manufacturing efficiency [64].

Tags can also be used to assist efficient recycling by providing accurate identification of objects in the recycling stream. For example, RFID tags are already allowed for marking lithium batteries under EU digital passport regulations. Tagging batteries may be useful in aiding efficient recovery of the valuable materials in the batteries. Tagged batteries may also improve safety in recycling, where lithium batteries, which are extremely flammable when they have significant residual charge [65], can start fires in recycling facilities [66].

### F. Digital Twins and RFID

The identification of individual objects will play an important role in the complete modeling of a (typically industrial) environment, where a digital twin of, for example, a warehouse, helps increase management efficiency and response speed. The digital twin concept links a physical object or system to a virtual copy that is continuously monitored, updated, and accessible to users [67]. RFID may play a key role for the development and accessibility of digital twins. RFID tags, for example, can serve as the physical links people can use to access virtual twins and relevant information related to the tagged objects. For the traditional retail sector, Ma et al. demonstrated the ability to read and localize tagged objects to generate a 3D real-time digital twin for a retail store [68]. With future integration of RFID readers on phones, RFID tags can be read by consumers to access information about the tagged objects, such as material composition, manufacturing history, and other information compliant with the EU's digital product passport initiative [69][70].

In addition to merely acting as the physical link to access the digital twin, RFID tags can play a broader role. For example, scanning tagged objects throughout their lifecycles allows for object traceability from production to disposal [69]. RFID tags integrated with low-power sensors can collect data that, with the tagged objects' identifiers, can be used to update digital twins. For example, Durgin et al. [71] proposed using low-power RFID and backscatter sensing motes to create digital spectrum twins. Cecchi et al. [72] demonstrated an indoor environment monitoring system that utilized a moving robot to read various RFID sensing tags in an environment, offering the possibility to create an environment digital twin. With RFID nodes being proposed for use in worker safety [73], agricultural monitoring [74], and vehicular tracking [75], there are opportunities to create digital twins that are enriched by sensing RFID nodes. However, as noted by Menon et al. [67], key challenges for digital twin creation and operation include data security, privacy, accuracy, and standardization of digital twin representations and the contributing data sources. These issues are discussed in Sections VI.B and VI.C below.

## V. THE LONG HAUL: THE NEXT SEVERAL DECADES

In this section we will discuss some more dramatic changes in the radio world that might occur over the course of the next several decades.

### A. Regulatory Reform and Artificial Intelligence

The current approach to regulating and using radio waves dates to the beginning of the 20th century, with minor elaborations over the last several decades. With the advent of artificial intelligence, and the promulgation of technologies generally targeted for short- and medium-range communications in developed environments, the (command-and-control with constrained-unlicensed) model may finally become obsolete.

Localized adaptive regulatory control based on artificial intelligence and ubiquitous detection could enable a broader use of spectrum for identification and characterization of the surrounding world, as well as general networking and navigation [76]. The existing system is a complex mix of administrative allocation, auctions, and unlicensed operation [77]. Overall reform based on global adaptive real-time allocation seems very challenging, but local allocation may be much more practical and set the stage for broader change. For example, Jeon et al. provide an example of how local base stations from multiple networks and/or vendors might cooperate in real time to optimize spectrum use, based on timeslots assigned by a local central coordinator [78]. Intelligent, software-driven networks that grant spectrum access, allocation, and sharing amongst different users are being researched with the development of specialized wireless test-beds, such as the Colosseum Project [79], [80], [81],and radio dynamic zones, such as POWDER-RDZ [82]. While this example is provided in the context of cellular communications, the timeslot-allocation approach is generalizable to a variety of radio interfaces, including passive RFID readers. (On the other hand, active beacon devices are somewhat more challenging to manage in this fashion, since we add a burden of time allocation signaling on to a power-constrained remote unit that would prefer to do nothing but send its periodic beacon.) It is possible that the propagation of such locally optimized sharing approaches, supported by adaptive machine-learning-based controls, may result in a new bottom-up radio access model, which can then be adapted and expanded by the regulatory bodies.

### B. Augmented Virtual Reality and RFID

The information obtained from an active or passive beacon associated with an object should become a part of what is perceived by a human using augmented reality, with sophisticated software to ascertain different details of the object, such as its type, history, and hazards. The merging of RFID with augmented reality is tantalizing for retail and warehouse environments to help consumers discover products, and workers to locate specific products.

Tavanti et al. [83] noted that the use of RFID with computer vision (CV) can overcome the line-of-sight requirement for CV detection. The RF sensing and identification of RFID tags provides additional information, in combination with the collected visual and motion information from a headset or smartphone, to help accurately detect and locate desired objects, creating an immersive digital twin with location depth that can be viewed as an augmented reality by the human user. Xiao et al. [84] noted that the fusion of RFID with augmented virtual reality has led to practical application use cases such as motion-tracking, target object localization, identification and information display, and access control. Li et al. [85] demonstrated and tested a mixed RFID and computer vision system that allows retail shoppers to use an augmented-reality (AR) headset, in real-time, to view relevant information for the product they are interacting with. Boroushaki et al. [86] prototyped an AR system that guides the human users to move towards and discover desired tagged object by conducting RF sensing and localization measurements with an AR headset that has a phased-array antenna on it.

The AR system does not have to be implemented with a headset. It can be integrated into a more common device such as a smartphone. Sato et al. [87] used a smartphone with an RFID reader to read multiple shelved tagged objects and map their locations on 3D models of indoor spaces that were generated from photos captured by the phone. This application even demonstrated an app on the smartphone that links desired RFID identifiers to their respective objects when operating in AR mode.

In most cases, tagged objects, even outside of retail or warehouse spaces, can be read by anyone with a reader anywhere at any time. As the number of tagged items increases, users with RFID readers in their personal devices can create rich augmented environments with the different tagged items being detected, located, and read by the readers. Each item's linked data from their digital twins would be accessible, easily visualized, and interconnected with each other, moving towards a delayed fulfillment of the Auto-ID Center's vision.

Many smartphones today have cameras and Bluetooth compatibility; thus, in the future, we can envision the fusion of RFID with barcodes, Bluetooth beacons, and other identification devices and methods to detect and identify anything anywhere at any time. The Auto-ID's lab vision of identifying everything might be achieved as the different methods of identification used by different products are connected and sensed by one device. With this fusion of sensing capabilities, a person's personal devices may generate an augmented reality from sensed data that will allow them to easily see and find everything in their vicinity and even guide them to the things they want to find. For example, retailers can in the future use all the identification information on a product and integrate it into an app, to help guide shoppers to find items of interest. And when these shoppers find the item they are looking for, they can then see the "story" behind the item, such as how it was manufactured and how it might be disposed of in the future. Augmented reality might serve as the endpoint of highly accessible and interactive digital twins in the physical world humans inhabit, encouraging users to explore history, attributes, and other relevant information of the objects in their inhabited environments.

The increased information depth of this augmented future can only be achieved if users trust the technology, and society can resolve issues with privacy and data management. What data and identifiers should people be allowed to see? Who should have access to and regulate the data generated?

## VI. Obstacles to Change

In his television series and book *Connections* [88], author James Burke examined the complex way in which real changes propagate through societies. Change in human society is often envisioned long before it becomes practical, due to the many barriers any innovation encounters, and the many-faceted web of knowledge and invention needed to surmount those obstacles. Changes impact the financial and social positions of existing stakeholders. Privileges that are accrued by individuals, corporate entities, or governments, may not be readily redistributed or shared.

Changes are also often impeded by the absence of incentives to change. For example, Heron of Alexandria demonstrated a steam-powered rotational device, capable of achieving up to 1500 RPM, in the first century AD [89]. Cylinders and pistons that would have been needed to extend the application of the engine were available, as were the first rail transport roads. However, the wide availability of enslaved human labor in the Alexandrian world probably limited the incentives to create means for moving goods. The dissemination of practical steam power in the 18$^{th}$ century was driven by the need for improved operation of pumps associated with mines, where human labor was challenged even when it was available [90].

In this section we examine some of the obstacles to the widespread implementation of radio-enabled knowledge we have discussed above.

### A. Barriers to Changes in Regulation

In the current regulatory environment, corporations or individuals have exclusive ownership of specific frequency bands in specific regions, typically under government-administered licenses. These organizations or individuals may perceive their interests in the existing regulatory system as valuable, and resist changes in regulation that would impact these privileges. The bottom-up regulatory reform approach envisioned in Section V.A above requires that at least local owners of spectral rights see a financial interest in some sort of real-time exchange arrangement, which is like to require some sort of market for those exchanges to take place.

In the United States, the Federal Communications Commission has used auctions to allocate many parts of the spectrum since 1994 [91]. For example, Auction 107, completed in 2021, involved total bids of US$81B for 5,684 licenses in the 3.7-4.2 GHz spectrum. Licensees are subject to complex requirements for services to be provided. Each licensee has commercial interests in the spectrum and geography purchased; corresponding government officials have an interest in administering the details of the system they have created. And there were 106 auctions before this one and several since it occurred. It is not trivial to see how these interests would be reconciled in a bottom-up adaptive

regulatory framework. Markets with no regulatory authority tend to be harmed by deceptive behavior and disputes between participants. So, coordinated development of mechanisms for exchanging usage rights under an overall regulatory authority would be needed. And the changes would need to eventually generalize to the numerous distinct regulatory domains across the commercial world.

*B. Barriers to Information Access*

The data associated with radio identification can be valuable to the creators of an object, who may resist sharing that data. For example, in the original vision of the Auto-ID Lab as described in Section II.B above, information about identified objects would be shared in a global database analogous to the World Wide Web. Anyone could read the identifier for an object and find out everything that was known about that object since it was manufactured. This scenario has not happened. Data is shared in very specific limited ways, for example between manufacturers and the retailers who sell their products.

The World Wide Web exists and works because it is a communications mechanism. It is in the interest of substantially all users to be able to reach other users. The disclosure of the corresponding IP address does not obligate the disclosure of any other private information about the holder of the address (although, of course, connections to the internet can expose users to malware if security precautions are not taken).

In contrast, exposing data about a specific object is not necessarily in the interest of the person or organization that made or sold the object. For example, a person or organization may purchase a medication in volume at low cost, with the pretense of selling to a large volume/high leverage medical customer, and then extract some of the items purchased to sell to smaller organizations or consumers at much higher prices, increasing their profit margin. The purchaser has no interest in exposing who got what item and when and where it was purchased. The end user of the medication may wish to know the chain of custody. Unless they are given an arbitrary right to have that information, they can only have some general assurance that manufacturing and shipment requirements have been met.

Similarly, if an individual wishes to purchase an item from someone, they might like to know when and where it was manufactured, how much it costs when new, and how it has been used by the current owner. Disclosure of that information is normal in some contexts, for example in the purchase of used cars (when the odometer reading has not been modified!), which are already uniquely identified. However, it is in the interest of the seller to represent an object as newer and initially more expensive than it is.

Finally, all of this must be justified by a demand for the product. If people cannot find the right sort of tools to employ the information that could be made available, or the applications of those tools that create sufficient value, there is no economic incentive to provide the data.

Data sharing can also impact the privacy of the owners of an object that has been sold or otherwise distributed, particularly in the context of reuse and recycling, where history benefits conflict with an individual's interest in not disclosing everything about their life.

Additionally, integrity of the data within traditional RFID networks have been of interest for security concerns, and current research has been looking into integrating blockchains and decentralized systems for improved traceability and authentication for RFID technology used in supply chains [92], [93].

*C. Dangers of Digitization of Human Perception*

The more we depend on supplementary nonlocal data for our existence in the world, the deeper the threat to that existence if the data is lost or compromised. For example, in today's world, an object on sale is usually labeled with a human-readable description of the object in addition to bar or QR codes and electromagnetic tags (see Figure 6 above). Objects on a retail shelf may be recognizable by a human irrespective of the label. However, in the envisioned virtual-reality world discussed in Section V.B above, actual human perception may be increasingly replaced by simulation. It is easy to imagine a progression in which the image of the real object is increasingly replaced by modeled images. This progression can contain tremendous utility in, for example, allowing the internal structure of the object to be seen when relevant, once information about an identified object becomes publicly available. But the use of simulated views of real objects also exposes the user to misrepresentation, distortion, and deception. Human history is filled with cases of false information about the world beyond that which is directly perceived by a given group of humans; virtual reality contains the hazard of extending that tradition of deception to the actual local world in which individual humans live, where until now understanding has been grounded in direct perception of physical objects.

## VII. CONCLUSION

Radio Frequency Identification is but one of many tools which continue to revolutionize the human perception of the world in which we live. With the introduction of readers into phones, digital twins in the cloud connected to physical objects, and augmented reality applications, the future for RFID promises exciting growth in the decades to come. To help these changes and growth to benefit the people that use them, the engineers and developers of these technologies need to plan for the consequent problems, as well as possible benefits, and communicate with stakeholders and the broader community.

ACKNOWLEDGMENT

The authors would like to acknowledge the assistance of Rich Marshall, Juho Partanen, and Jeffrey Dungen in the production of this paper. Thank you to all the past and current members of the Propagation Group at the Georgia Institute of Technology for their extensive thought and research about the next generation of RFID.